\documentstyle[psfig,times]{mn}

\def\etal{et~al.}
\def\spose#1{\hbox to 0pt{#1\hss}}
\def\lta{\mathrel{\spose{\lower 3pt\hbox{$\mathchar"218$}}
     \raise 2.0pt\hbox{$\mathchar"13C$}}}
\def\gta{\mathrel{\spose{\lower 3pt\hbox{$\mathchar"218$}}
     \raise 2.0pt\hbox{$\mathchar"13E$}}}
 
\title[Radio--loud AGN in the SDSS]{A sample of radio-loud AGN in the
Sloan Digital Sky Survey}

\author[P.~N.~Best \etal]{P.~N.~Best,$^1$\thanks{Email:
pnb@roe.ac.uk} G. Kauffmann$^2$, T. M. Heckman$^3$, \v{Z}. Ivezi{\'c}$^4$ \\
$^1$ Institute for Astronomy, Royal Observatory Edinburgh, Blackford Hill,
Edinburgh EH9 3HJ, UK\\
$^2$ Max-Planck-Institut f{\"u}r Astrophysik, Karl-Schwarzschild-Str. 1,
D-85748 Garching, Germany\\
$^3$ Department of Physics \& Astronomy, The Johns Hopkins University,
Baltimore, MD 21218, USA\\
$^4$ Princeton University Observatory, Peyton Hall, Princeton, NJ
08544-1001, USA}

\pagerange{\pageref{firstpage}--\pageref{lastpage}}
\pubyear{2004}
\begin{document}
\label{firstpage}

\maketitle

\begin{abstract}
\noindent A sample of 2712 radio--luminous galaxies is defined from the
second data release of the Sloan Digital Sky Survey (SDSS) by
cross--comparing the main spectroscopic galaxy sample with two radio
surveys: the NRAO VLA Sky Survey (NVSS) and the Faint Images of the Radio
Sky at Twenty centimetres (FIRST) survey. The comparison is carried out in
a multi--stage process and makes optimal use of both radio surveys by
exploiting the sensitivity of the NVSS to extended and multi--component
radio sources in addition to the high angular resolution of the FIRST
images.  A radio source sample with 95\% completeness and 98.9\%
reliability is achieved, far better than would be possible for this sample
if only one of the surveys was used. The radio source sample is then
divided into two classes: radio--loud AGN and galaxies in which the radio
emission is dominated by star formation. The division is based on the
location of a galaxy in the plane of 4000\AA\ break strength versus radio
luminosity per unit stellar mass and provides a sample of 2215 radio--loud
AGN and 497 star forming galaxies brighter than 5\,mJy at 1.4\,GHz. A full
catalogue of positions and radio properties is provided for these sources.
The local radio luminosity function is then derived both for radio--loud
AGN and for star-forming galaxies and is found to be in agreement with
previous studies.  By using the radio to far-IR correlation, the radio
luminosity function of star forming galaxies is also compared to the
luminosity function derived in the far--infrared. It is found to agree
well at high luminosities but less so at lower luminosities, confirming
that the linearity of the radio to far-IR correlation breaks down below
about $10^{22}$W\,Hz$^{-1}$ at 1.4\,GHz.
\end{abstract}

\begin{keywords}
galaxies: active --- galaxies: evolution --- galaxies: luminosity function
--- galaxies: starburst --- radio continuum: galaxies --- surveys
\end{keywords}

\section{Introduction}

In recent years, new radio surveys such as the National Radio Astronomy
Observatory (NRAO) Very Large Array (VLA) Sky Survey (NVSS; Condon \etal\
1998)\nocite{con98} and the Faint Images of the Radio Sky at Twenty
centimetres (FIRST) survey \cite{bec95} have covered substantial fractions
of the sky down to milli-Jansky flux densities, at vastly higher angular
resolution than their predecessors. Such radio surveys are dramatically
advancing our understanding of extragalactic radio sources, by permitting
detailed statistical studies to be carried out. In order to reap the full
benefit of these surveys, it is necessary to optically identify the radio
sources, so as to obtain spectroscopic redshifts and determine the
properties of their host galaxies. The availability of new large galaxy
redshift surveys, especially the 2-degree Field Galaxy Redshift Survey
(2dFGRS; Colless \etal\ 2001)\nocite{col01} and the Sloan Digital Sky
Survey (SDSS; York \etal\ 2000; Stoughton \etal\
2002)\nocite{yor00,sto02}, means that optical identifications and
redshifts are available for large samples of nearby radio sources and
allows comprehensive statistical analyses of their host galaxy properties
to be carried out.

Automated cross--correlation of surveys across different wavelength
regimes has a long history in astronomy. It is important that this process
should maximize both the completeness and the reliability of the resulting
sample, so considerable care needs to be taken in choosing the parameters
that determine whether objects in different catalogues are indeed
associated.  In the case of optical identification of radio sources, the
choice of radio survey is important. This is because many radio sources
are extended, with sizes from a few arcsec up to tens of arcmins, and in
high angular resolution surveys different components of the same source
may be resolved into distinct sources. Surveys at lower angular resolution
detect most sources as single components, and also have good sensitivity
to extended radio structures, but the high surface density of possible
optical counterparts (over 4000 per square degree at high Galactic
latitudes in the Palomar Observatory Sky Survey) limits the reliability of
the optical matching.

The first radio sky surveys were carried out at very low angular
resolution and detected only the brightest radio sources. The resolution
of these surveys was too low to allow identification of the host galaxies
without detailed radio follow-up observations of the detected sources;
this was time-consuming and meant that only small samples of galaxies
could be studied (see discussion in McMahon \etal\ 2002).\nocite{mcm02}
The NVSS was the first radio survey of sufficiently high angular
resolution (45 arcsec) to permit automated cross--correlation with optical
surveys. Machalski \& Condon \shortcite{mac99} cross--correlated the NVSS
with the Las Campanas Redshift Survey (LCRS; Shectman \etal\
1996)\nocite{she96}, identifying 1157 radio--emitting galaxies. Machalski
\& Godlowski \shortcite{mac00} used this sample to derive the local radio
luminosity function. Using far--infrared data available for the LCRS they
were also able to separate the luminosity function into a radio--loud
active galactic nuclei (AGN) component, which dominates at high radio
luminosities, and a lower-luminosity component due to star--forming
galaxies that emit in the radio predominantly due to the synchrotron
emission from supernova remnants.  Similarly, Sadler \etal\
\shortcite{sad02} cross-correlated the NVSS with galaxies from the first
data release of the 2dFGRS, defining a sample of 912 radio sources which
form a basis for further detailed studies (e.g. Best 2004).\nocite{bes04a}

The 45 arcsec resolution of the NVSS has the advantage of being
sufficiently large that $\sim 99$\% of radio sources are contained within
a single NVSS component. With the exception of a few very large sources,
the NVSS is also able to detect the entirety of the radio
emission. However, the poor angular resolution of the NVSS leads to
significant uncertainties in cross-identifying the radio sources with
their optical host galaxies and there is a trade-off between the
reliability of the matched sample and its completeness. Sadler \etal\
\shortcite{sad02} accepted radio sources within a matching radius of 10
arcseconds from an optical galaxy, leading to a catalogue that was $\sim
90$\% complete, but in which 5--10\% of the matches are expected to be
false identifications.

Samples with much higher reliability can be derived using the FIRST
catalogue, due to its superior angular resolution ($\sim 5$
arcsec). Ivezi{\'c} \etal\ \shortcite{ive02} cross--correlated the FIRST
survey with the SDSS imaging sample. Under the assumption that all true
identifications of point radio sources would have radio--optical
positional offsets of less than 3 arcsec, they concluded that the optimal
matching radius for cross--correlation was 1.5 arcsec, for which they
derived a completeness for radio point sources of 85\% and a contamination
rate of only 3\%.

However, at the high angular resolution of FIRST, new problems arise.
FIRST is not sensitive to extended radio structures because of a lack of
short antennae baselines, and resolves out the extended emission of radio
sources. As a result, the total radio luminosity of sources that are
larger than a few arcseconds will be systematically low (cf. Becker \etal\
1995).\nocite{bec95} In extreme cases, some larger radio sources are
missed. These effects introduce systematic biases into the derived radio
source sample. In addition, many extended radio sources are split into
multiple components by FIRST. Matching routines therefore need to be
developed to account for the possible multi-component nature of radio
sources.

The first attempt to automate such a routine was by Magliocchetti \etal\
\shortcite{mag98b}, who used a `collapsing algorithm' to identify
multi--component FIRST sources. They considered all pairs of sources with
separations below 3 arcmins, and merged into a single combined source all
pairs with separations below $100 \left( S_{\rm tot} / 100{\rm mJy}
\right)^{0.5}$ arcsec and flux densities within a factor of four of each
other.  This method is simple and works well for classical double--lobed
radio sources, but accounts poorly for core--jet sources or sources with
large asymmetries.

Ivezi{\'c} \etal\ \shortcite{ive02} improved on this by first
cross-correlating all FIRST sources with the SDSS (thereby picking up all
sources with a core component) and then adding candidate double--lobed
radio sources to this sample. These were identified by comparing the
mid-points of all FIRST pairs with separations below 90 arcsec with the
galaxies in the optical catalogue, and accepting all matches with offsets
below 3 arcsec. They estimated that such double sources contribute less
than 10\% of all radio sources.

McMahon \etal\ \shortcite{mcm02} carried out a detailed study of the
properties of multi--component FIRST sources by comparing isolated pairs
of FIRST sources with optical Automated Plate Measuring Machine (APM)
scans of the Palomar Observatory Sky Survey (POSS) plates.  For core--jet
type sources, where the optical counterpart is associated with one of the
radio components, they found that the radio components usually have very
different flux densities and that the component with the optical
counterpart is usually brighter and is frequently unresolved in the
radio. In contrast, if the optical counterpart is located between the two
radio components, the two radio components usually have comparable flux
densities and similar radio sizes (ie. both are consistent with being
radio lobes, not one unresolved core and an extended radio lobe). In this
case, the optically identified galaxy is typically located fairly close to
the flux-weighted mean position of the two radio components. This
information is extremely useful in the identification of multi--component
FIRST sources.

Because the main spectroscopic galaxy sample of the SDSS has rather low
median redshift ($z \sim 0.1$), the problems described above associated
with identifying extended radio sources will be more severe.  This paper
thus presents a hybrid method, using information from both NVSS and FIRST
in order to take advantage of the strong points of both surveys and avoid
the systematic errors that arise in using only one of them.  The layout of
the paper is as follows. In Section~\ref{radoptsamps}, the salient points
of the SDSS, NVSS and FIRST surveys are summarised.
Section~\ref{radmatch} then discusses the cross-matching of these surveys
to identify the radio source sample. Section~\ref{radloudagn} describes
how true radio-loud AGN are separated from sources where the radio
emission is associated with star formation activity. The local radio
luminosity functions of radio-loud AGN and star--forming galaxies are
derived in Section~\ref{radlumfunc}, and the radio luminosity function of
star--forming galaxies is compared to that derived at far-infrared
wavelengths. Conclusions are drawn in Section~\ref{discuss}.  In an
accompanying paper \cite{bes05b} the host galaxies of the radio--loud AGN
are investigated in detail.  Throughout the paper, the values adopted for
the cosmological parameters are $\Omega_m = 0.3$, $\Omega_{\Lambda} =
0.7$, and $H_0 = 70$\,km\,s$^{-1}$Mpc$^{-1}$.

\section{The radio and optical galaxy samples}
\label{radoptsamps}

\subsection{The SDSS Spectroscopic Sample}
\label{sdsssamp}

The Sloan Digital Sky Survey (York \etal\ 2000; Stoughton \etal\ 2002, and
references therein)\nocite{yor00,sto02} is an optical imaging ({\it
u,g,r,i,z} bands) and spectroscopic survey of about a quarter of the
extragalactic sky, being carried out at the Apache Point Observatory.  The
spectroscopic sample considered in this paper is a sample of about 212,000
objects with magnitudes $14.5 < r < 17.77$, spectroscopically confirmed to
be galaxies, drawn from the `main galaxy catalogue' of the second data
release (DR2) of the SDSS. This sample of galaxies is described by
Brinchmann \etal\
\shortcite{bri04b}. The galaxies have a median redshift of $z \sim 0.1$.

The SDSS spectra cover an observed wavelength range of 3800 to 9200\AA, at
an instrumental velocity resolution of about 65km\,s$^{-1}$. The spectra
are obtained through 3-arcsec diameter fibres, which corresponds to 5.7
kpc at a redshift of 0.1; at this redshift the spectra therefore represent
a large proportion (up to 50\%) of the total galaxy light, while for the
very lowest redshift objects they are more dominated by the nuclear
emission.

As described in Brinchmann \etal\ \shortcite{bri04b}, a variety of
physical parameters for these galaxies have been derived from the
photometric and spectroscopic data, and catalogues of these parameters are
publically available.  These include total stellar masses, mass-to-light
ratios, 4000\AA\ break strengths, H$\delta$ absorption measurements and
estimates of dust attenuation \cite{kau03a,kau03b}; accurate emission line
fluxes, after subtraction of the modelled stellar continuum to account for
underlying stellar absorption features (Kauffmann \etal\
2003c,\nocite{kau03c} Tremonti \etal\, in preparation); galaxy
metallicities \cite{tre04}; parameters measuring optical AGN activity,
such as emission line ratios, and galaxy velocity dispersions (hence black
hole mass estimates; Kauffmann \etal\ 2003c, Heckman \etal\
2004).\nocite{kau03c,hec04} These parameters have been adopted for the
analyses of this paper: the reader is referred to the papers referenced
above for detailed information about the methods used to derive them.

It should be emphasised that the use of the SDSS main galaxy catalogue as
the basis sample for this study means that objects classified as `quasars'
by the automated SDSS classification pipeline (Schlegel \etal\, in
preparation) are excluded. These objects are excluded because of the
influence of the direct non--stellar continuum light from the active
nucleus. This affects the observed optical magnitudes, preventing clean
magnitude--limited samples from being derived, and prohibits the host
galaxy parameters discussed above from being accurately determined. The
number of AGN excluded in this way is very small ($\lta 3$\% out to
$z=0.1$\footnote{A search of the SDSS DR2 database reveals only 393
objects classified as `quasars' in the redshift range $0.03 < z < 0.1$,
compared to 16661 objects classified as emission--line AGN by Kauffmann
\etal\ \shortcite{kau03c}.}), since this exclusion only applies to the
most luminous Type-I AGN: the SDSS pipeline classifies most low luminosity
Type-I AGN as `galaxies' rather than `quasars', and so they are
retained. Kauffmann \etal\ \shortcite{kau03c} estimate that 8\% of AGN in
the main galaxy sample have broad emission lines, and therefore are
strictly Type-Is. They also demonstrate that for these low luminosity
objects the non-stellar continuum light has a negligible effect on the
physical parameters derived for the host galaxy.

\subsection{The NVSS and FIRST Radio Surveys}

The NVSS \cite{con98} and FIRST \cite{bec95} surveys are two radio surveys
that have been carried out in recent years using the VLA radio synthesis
telescope at a frequency of 1.4\,GHz. The NVSS was observed with the array
in D-configuration (DnC configuration for the most southerly fields),
which provides an angular resolution of 45 arcseconds. This survey covers
the entirety of the sky north of $-40^{\circ}$ declination, down to a
limiting point source flux density of about 2.5\,mJy. The FIRST
observations were carried out in B-array configuration, which provides a
much higher angular resolution of $\sim 5$ arcsec. This survey was
designed to study the region of sky that will be observed by the SDSS, and
therefore overlaps with this very closely. It reaches a limiting flux
density of about 1\,mJy for point sources.

\section{Cross-matching of the SDSS spectroscopic sample using a
combination of NVSS and FIRST}
\label{radmatch}

The FIRST and NVSS surveys are highly complementary for identifying radio
sources associated with nearby galaxies; NVSS provides the sensitivity to
large--scale radio structures required to detect all of the emission from
extended radio sources, while FIRST provides the high angular resolution
required to reliably identify the host galaxy. In order to identify radio
sources associated with galaxies in the SDSS spectroscopic sample, a
hybrid method using both radio surveys has been derived. A broad overview
of the steps in this process is as follows:

\noindent 1. SDSS galaxies lying outside the sky coverage of the FIRST
survey were excluded. Galaxies close to very bright radio sources were
also excluded, because the noise of the NVSS images is much greater in
these regions. Finally, galaxies with redshifts below 0.01 were excluded,
because at these low redshifts the galaxies are very extended and their 
optical positions are consequently uncertain.

\noindent 2. The remaining sample was cross--correlated with the NVSS
catalogue. A list of candidate galaxies that might be associated with
multi-NVSS-component radio sources was derived.

\noindent 3. These multi--NVSS--component candidates were investigated; by
necessity, a small proportion of this analysis had to be done visually
rather than through automated procedures. If a galaxy was confirmed to be
associated with a multi--NVSS--component source, the integrated flux
densities of the NVSS components were summed to provide the radio source
flux density.

\noindent 4. All galaxies matched with a single NVSS source were then
cross--correlated with the FIRST catalogue. Note, however, that the
presence of a FIRST counterpart was not {\em required} for a source to be
accepted.  If there was no FIRST counterpart, then the source was accepted
or rejected solely upon its NVSS properties.

\noindent 5. If a single FIRST counterpart was associated with the NVSS
source, then the source was accepted or rejected on the basis of the
properties of the FIRST counterpart. For accepted matches, however, the
adopted radio flux density was taken from the NVSS data.

\noindent 6. If multiple FIRST components were associated with the NVSS
source, then the source was accepted if it satisfied criteria for a
single-component source (with unrelated additional FIRST sources) or for a
radio source with multiple FIRST components.  Again, the NVSS catalogue
was used to provide the most accurate measure of the radio flux density.

The exact criteria for accepting and rejecting matches in the procedures
outlined above were tested and refined using Monte--Carlo simulations. Ten
catalogues of random sky locations were constructed, over the same sky
area as the SDSS survey. Each catalogue contained the same number as
positions as the list of SDSS galaxies, and these random catalogues were
taken through exactly the same steps of cross--comparison with the radio
data as the SDSS galaxy catalogue. In the subsections that follow, the
resulting optimal selection criteria are described, together with the
completeness and reliability estimates provided by the Monte--Carlo
simulations.

Note that the flux densities adopted for the NVSS sources are true
integrated flux densities, rather than the peak flux densities quoted in
the NVSS catalogues. The formulae for conversion of peak flux densities to
integrated flux densities are provided by Condon \etal\
\shortcite{con98}. Only those radio sources with total flux densities
(after summing NVSS components if necessary) above 5\,mJy are
retained. This flux density limit corresponds to approximately 10 times
the noise level of the NVSS maps, and is adopted because: (i) at this
significance level, all sources should be real and have well--determined
positions; (ii) at this flux density limit, the sample is as sensitive to
extended single--component NVSS sources (which will have a lower peak flux
density) as it is to point sources, and the sensitivity to
multi--component NVSS sources will not be significantly worse (for
example, a 5\,mJy source composed of two individual components of 2.5\,mJy
would be found). The 5\,mJy limit corresponds to about
$10^{23}$W\,Hz$^{-1}$ at redshift $z \sim 0.1$, which is approximately
where the radio luminosity function switches from being dominated by star
forming galaxies (low luminosities) to being dominated by AGN (high
luminosities).

\begin{figure*}
\centerline{
\psfig{file=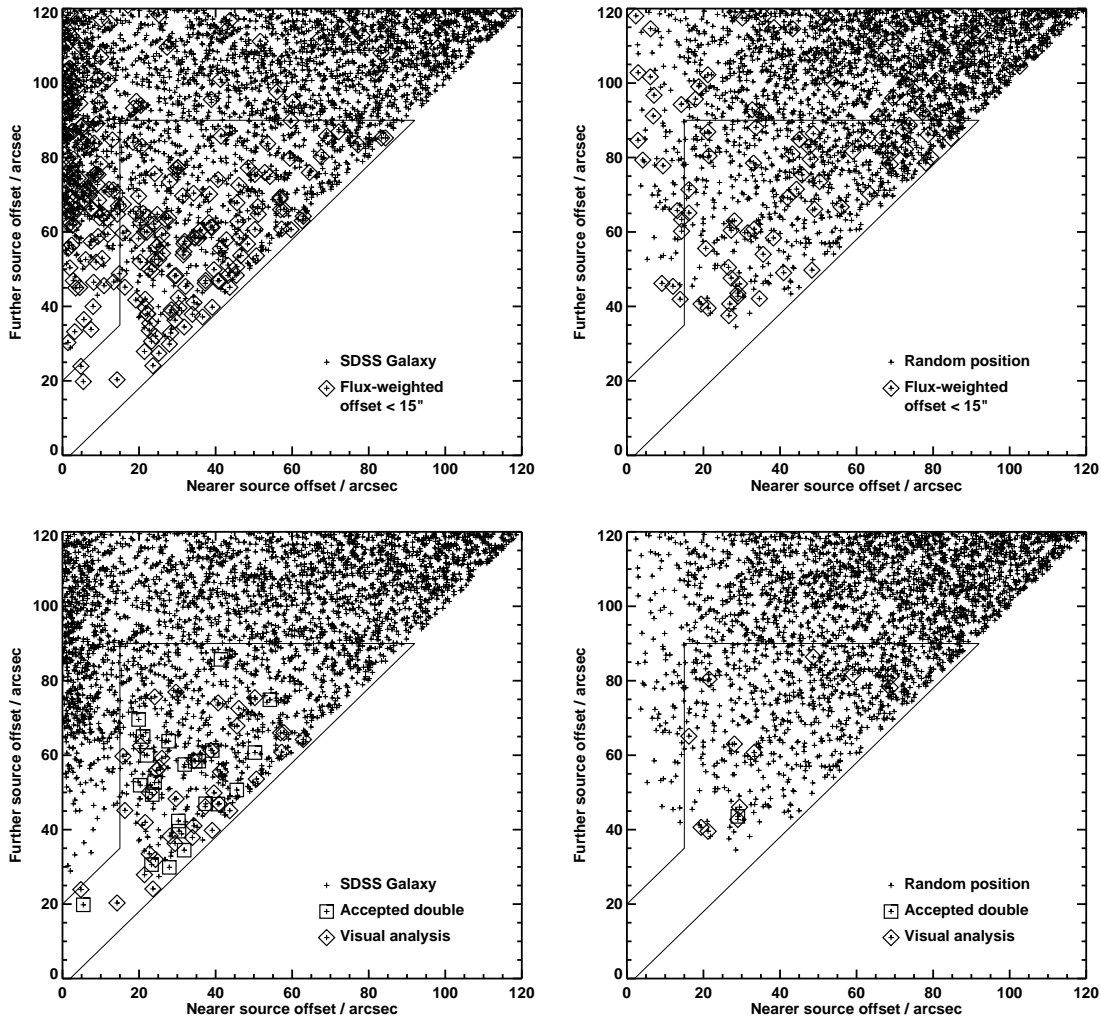,width=15cm,clip=} 
}
\caption{\label{nvssdbls} For those SDSS galaxies (left) and an equivalent
number of random positions (right) which have two NVSS sources within 3
arcmins, the plots show the positional offsets of the two NVSS sources
from the optical galaxy or random position. In the upper panels, the
diamonds indicate those cases where the flux--weighted mean position of
the two NVSS sources is within 15 arcsec of the optical/random
position. The solid lines enclose the region in which large numbers of
candidate double sources are found, without being swamped by false
identifications. The lower plots then show the result of comparing these
candidate doubles with the FIRST catalogue. The criteria for the
candidates to be accepted, rejected, or referred to visual analysis are
described in the text: this leads to a number of accepted doubles (with
very few false detections), and a small sample of sources to be visually
analysed, of which of order half will turn out to be genuine.}
\end{figure*}

\subsection{Identification of multi-component NVSS sources}

In order to search for possible multi-component NVSS sources, a search was
made for multiple sources within a radius of 3 arcmins from each optical
galaxy.  This distance was selected to be large enough that any genuine
multi-component radio source should have at least two matches, but still
much smaller than the typical separation of NVSS sources (8-10 arcmins).
\smallskip

\noindent{\bf Candidate NVSS doubles.} For galaxies with two NVSS matches
within 3 arcmins, the top panels of Fig~\ref{nvssdbls} compare the offsets
of the two NVSS matches from the optical position for SDSS galaxies (left)
and for an equivalent number of random positions (right). There are a
large number of SDSS galaxies for which the nearer NVSS component lies
within 15 arcsec of the optical galaxy; these are predominantly galaxies
containing a single--component NVSS source and the other NVSS source is
physically unrelated.  Such sources were classified as single--component
matches (see below).

In addition to these, there is a clear excess of SDSS galaxies (compared
to random) that have the two NVSS components each offset by 20-50 arcsec
from the optical position. For these systems, the flux-weighted mean
position of the two NVSS sources is often within 15 arcsec of the optical
galaxy (indicated by the diamonds in the upper panel of
Fig~\ref{nvssdbls}). Candidate NVSS doubles are therefore selected to be
sources with both NVSS components closer than 90 arcsec, a flux-weighted
mean position closer than 15 arcsec, and the nearer component offset by
more than whichever is smaller out of 15 arcsec and the offset of the
second source minus 20 arcsec. These selection criteria are indicated by
the lines on Fig~\ref{nvssdbls}.  The 90 arcsec limit is chosen since
larger offsets are relatively unlikely and the contamination by random
galaxies gets increasingly large beyond this. Even with this limit, there
is still significant contamination, but the next step of comparison with
FIRST helps to alleviate much of this.

\begin{figure*}
\centerline{
\psfig{file=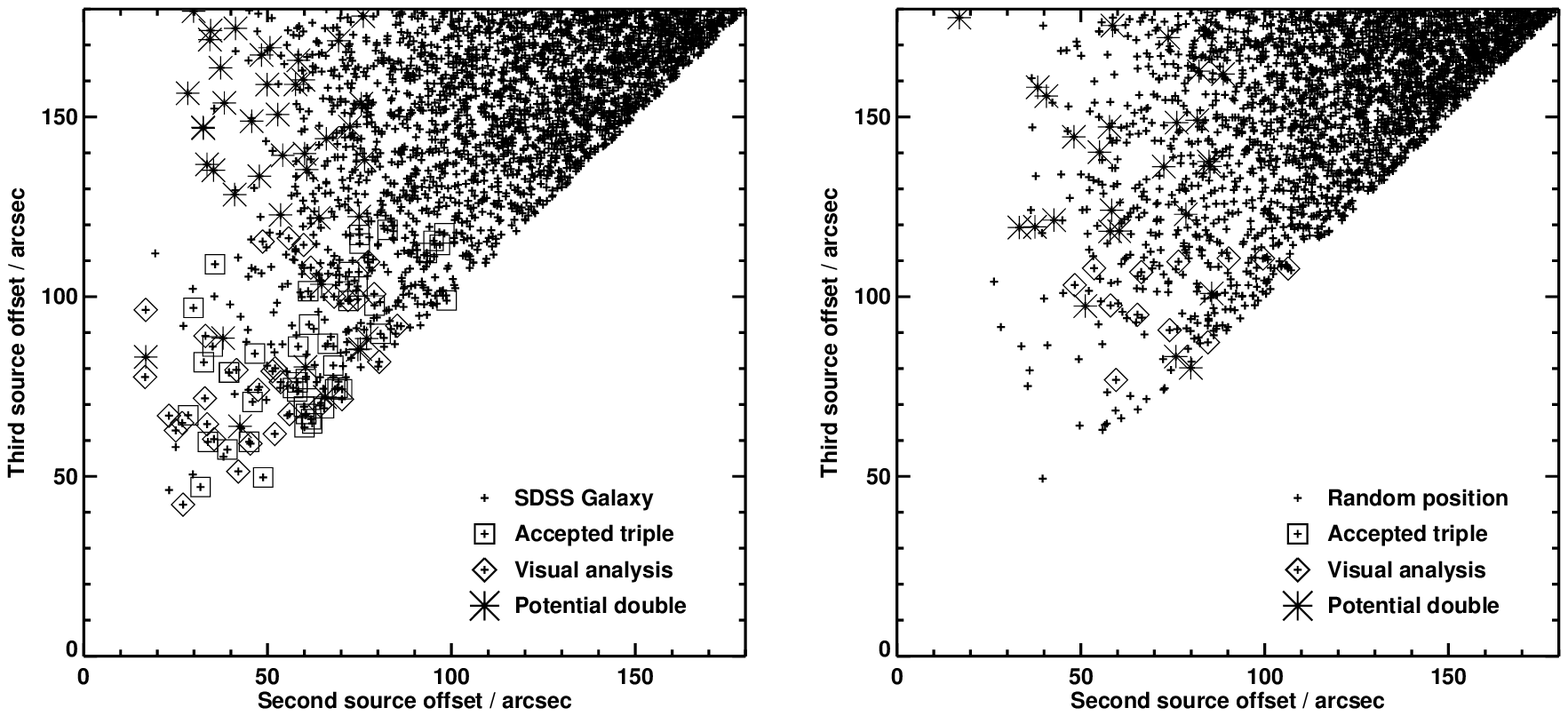,width=15cm,clip=}
}
\caption{\label{nvsstrpls} The positional offsets of the second and third
nearest NVSS sources from those SDSS galaxies (left) and an equivalent
number of random positions (right) which have three NVSS sources within 3
arcmins. As for Figure~\ref{nvssdbls}, the criteria for the candidates to
be accepted, rejected, classified as potential doubles (on the basis of
their NVSS properties -- further investigation with FIRST is then carried
out to confirm or reject their double status), or referred to visual
analysis are described in the text.}
\end{figure*}

All of these candidate doubles were cross-correlated with the FIRST
catalogue. If these are true extended doubles then they may have a central
FIRST component associated with a radio source core, and in addition they
are likely to be missing flux in the FIRST data due to their extended
nature; indeed they may well be undetected by FIRST. If they are not true
doubles, but two individual NVSS sources, then it is likely that a single
or double FIRST counterpart is present at each NVSS location, with little
missing flux.  The candidate doubles were thus classified into three
categories:

\noindent (a) accepted doubles: sources were accepted as NVSS doubles if
they either have a FIRST source within 3 arcsec of the optical position,
or they satisfy the following three conditions  (i) no detected FIRST component
(ie. all of the flux is resolved out by FIRST); (ii) both NVSS components
lie within 60 arcsec of the SDSS position (larger sources may
have additional NVSS components outside of the 3 arcmin limit, and so need
to be checked visually); and (iii) the angle NVSS1-SDSS-NVSS2 greater than
135 degrees (ie. consistent with a double radio source with a bend of
$<45$ degrees).

\noindent (b) rejected doubles: those sources with 3 or fewer FIRST
components, all further than 15 arcsec from the optical galaxy, and with
total flux greater than half of the sum of the two NVSS fluxes, were
rejected.

\noindent (c) uncertain cases: any sources not satisfying either of the
above conditions were classified as uncertain, and referred for visual
analysis.

The lower two plots of Figure~\ref{nvssdbls} show the results of this
classification of candidate doubles for the SDSS sources and an equivalent
number of random positions. 
\smallskip

\noindent{\bf Candidate NVSS triples.} Galaxies with 3 NVSS components
within 3 arcmins could represent one of four possibilities: (i) a triple
radio source associated with the galaxy; (ii) a double radio source
associated with the galaxy, together with an unassociated NVSS source;
(iii) a single radio source, with two unassociated sources (or an
unassociated double source); (iv) 3 unassociated NVSS sources. It is the
first two possibilities that are the concern for the multiple--source
analysis.

Comparison between the SDSS galaxies and the random positions
(Fig~\ref{nvsstrpls}) suggests that a source should be classified as a
potential triple if all three components are within 120 arcsec, and one of
the following three conditions is also satisfied: (i) the flux weighted
mean position of all three components is within 15 arcsec of the optical
galaxy position; (ii) the flux weighted mean position of the two more
distant components is within 15 arcsec of the optical galaxy position
[this for the case where these are the two outer lobes of a radio source,
and the nearest component is a feature in the jet of one of the sources];
(iii) the nearest component is within 15 arcsec of the optical galaxy,
with the second and third components both within 90 arcsec and the angle
NVSS2-NVSS1-NVSS3 greater than 135 degrees [this is the case where the
nearest component corresponds to the core of the triple: the offset and
angle classification requirements distinguish this from a single component
source with two unassociated sources].

Galaxies which satisfied these constraints were investigated using FIRST
to accept or reject obvious cases.  Galaxies were accepted as triples if
they possessed a FIRST source within 3 arcsec of the optical galaxy
position. They were rejected if, as for the doubles, they had 3 or fewer
FIRST components, all further than 15 arcsec from the optical galaxy, with
total flux equal to at least half of the sum of the three NVSS fluxes. The
remainder of the galaxies were referred to visual
analysis. Figure~\ref{nvsstrpls} compares the results of this analysis for
the SDSS galaxies and the random sample.

Galaxies that were neither classified as triples nor visually inspected
were then investigated to see if they were associated with a double radio
source. Each of the three NVSS pairs was checked using the double source
analysis described above.
\smallskip

\noindent{\bf Candidates with 4 or more NVSS matches.} Galaxies with 4 or
more NVSS sources within 3 arcminutes of the galaxy are likely either to
be the host galaxy of a multiple-component radio source, or to lie close
to one.  No such cases were accepted without visual analysis.  Visual
analysis was carried out on all galaxies with 5 or more NVSS matches, as
well as on galaxies with 4 NVSS matches where either the mean position or
the flux-weighted mean position of the 4 NVSS sources was within 30 arcsec
of the optical galaxy.  All other galaxies were not considered to be
quadrupoles, but were examined for potential triples and doubles using the
criteria described above.

Overall, as a result of the multiple source analysis, 60 SDSS galaxies
were accepted as NVSS multiple-component sources, compared to only 0.3
multiple-component sources predicted from the Monte--Carlo simulations
using random positions. This corresponds to less than 1\% contamination. A
further 277 sources (0.13\% of the original sample) required visual
analysis, of which 109 were ultimately accepted as genuine sources. This
total of 169 accepted multi--NVSS--component sources corresponds to about
6\% of the entire SDSS radio source sample.

\begin{figure}
\centerline{
\psfig{file=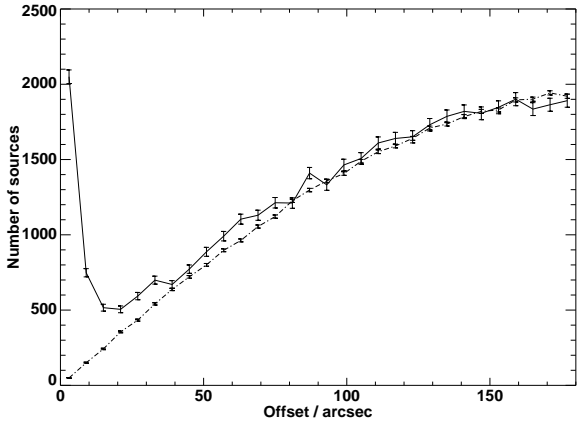,angle=90,width=8cm,clip=}
}
\caption{\label{nvssoffs} The positional offsets between SDSS galaxies and
their nearest NVSS source (solid line) compared to the expectation derived
from random positions (dot--dash line). There is a clear excess at small
offsets, associated with true matches. Few if any true matches are
expected above a separation of 15 arcsec; the excess at these radii is
caused by the clustering of optical galaxies.}
\end{figure}

\subsection{Single-component NVSS matches}

For all galaxies not classified as multiple NVSS sources,
Figure~\ref{nvssoffs} compares the distribution of offsets between SDSS
galaxies and their nearest NVSS source with the result obtained for random
positions.  There is a clear excess of sources associated with SDSS
galaxies at small radii. Because multiple sources have been removed and
the analysis is restricted to brighter sources with well-defined
positions, true sources with offsets larger than about 15 arcsec are not
expected, but the excess is significant out to at least 100 arcsec. At
these large radii the excess is not due to true associations, but rather
is the result of the clustering of optical galaxies: on average there are
more galaxies within $\sim 100$ arcsec of an optical galaxy than within
the same distance of a random position, and hence there is an increased
chance of finding an unassociated radio source. Note that in principle
this effect could be accounted for by incorporating an appropriate
correlation function into the positions of objects in the random
catalogues; however, full knowledge of the environments of radio source
hosts would be required to do this properly.

Integrating under the two curves of Figure~\ref{nvssoffs} out to a
separation of 15 arcsec gives 2973 matches for the SDSS galaxies and 311
random matches, so the NVSS data alone would suffer from $\approx 10$\%
contamination if a 15 arcsec separation were adopted. This falls to $\sim
6$\% contamination at 10 arcsec, but at a cost of reducing the
completeness by 10\%. The completeness can be improved by including
information from the FIRST data.  All SDSS galaxies with a
single-component NVSS match within 30 arcsec are thus cross-correlated
with the FIRST catalogue to determine the number of FIRST components
within the 30 arcsec radius.

\begin{figure}
\centerline{
\psfig{file=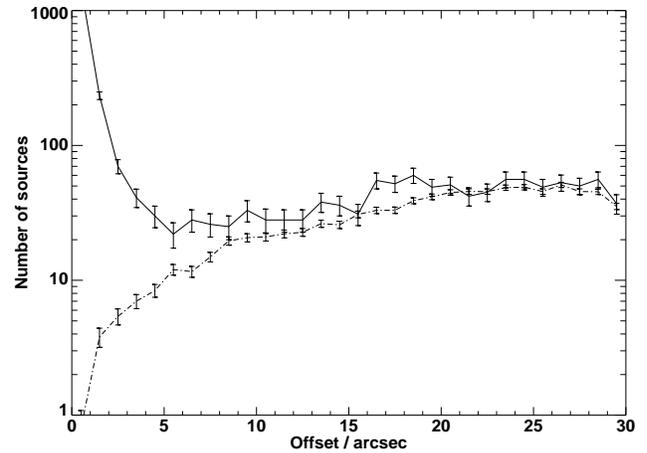,angle=90,width=8.6cm,clip=}
}
\caption{\label{firstoffs} The positional offsets between SDSS galaxies
and the FIRST source (solid line), for those SDSS galaxies with a single
NVSS match of $>5$ mJy flux density within 30 arcsec and a single FIRST
component associated with that. This is compared to the expectation
derived from random positions (dot--dash line). The excess over the 10--30
arcsec range is likely associated with the clustering of optical galaxies;
most of those with offsets less than 10 arcsec are likely to be true
identifications.}
\end{figure}

\begin{figure*}
\centerline{
\psfig{file=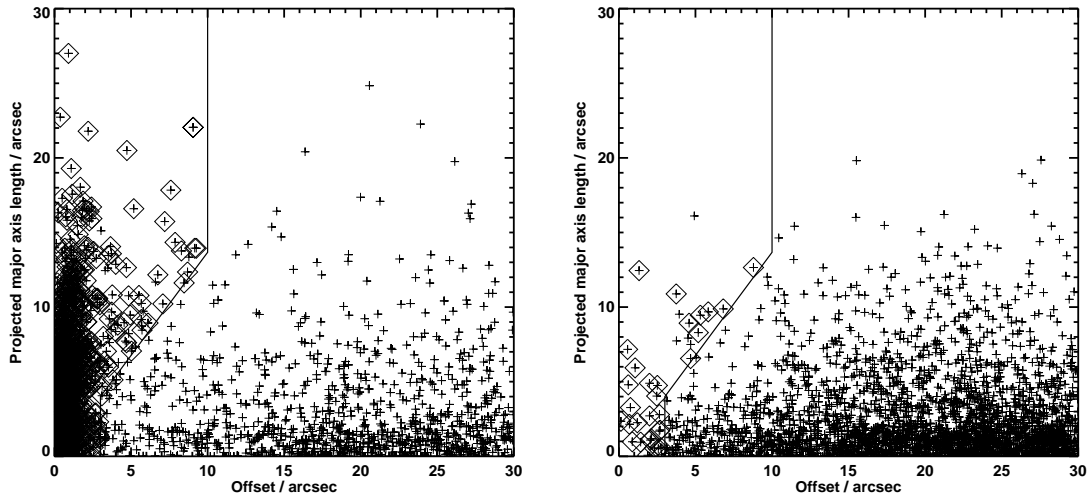,width=15cm,clip=}
}
\caption{\label{firstoffsize} Left: for those SDSS galaxies with a single
associated FIRST source, the projected major axis size of the FIRST source
is plotted against its positional offset from the optical position. Right:
the equivalent plot for the random positions. It is clear that much of the
excess of SDSS-FIRST offsets of 3--10 arcsec is associated with extended
FIRST sources. The solid lines indicate the allowed limits to the offset
for a source to be considered a possible match, namely that the offset
must be either less than 3 arcsec or be both between 3 and 10 arcsec and
less than 75\% of the projected major axis length. The diamonds indicate
the finally accepted sources: those with offsets less than 3 arcsec or
which lie to the left of the solid line and are oriented within 30 degrees
of the offset vector (see text for full details).}
\end{figure*}

\subsubsection{Sources with no FIRST matches}

These sources are either variable radio-loud AGN which have faded between
the NVSS and FIRST observations or they are extended radio sources which
are resolved out of the FIRST dataset. In either case they should be
retained if they are associated with the optical galaxy. These NVSS
sources are accepted as matches if they lie within 10 arcsec of the
optical position. 134 sources are retained in this way (compared to 8.7
random); this corresponds to approximately 5\% of the total radio source
sample.

\subsubsection{Sources with one FIRST match}

For cases with one FIRST source within 30 arcsec, Figure~\ref{firstoffs}
compares the distribution of separations of the FIRST source and the SDSS
galaxy with the result obtained for the random catalogue.  An excess of
FIRST sources with respect to random is seen at all separations; at large
separations this is the result of the clustering of optical galaxies, as
discussed above. The excess becomes particularly pronounced at separations
less than 10 arcsec, but at separations larger than $\sim 3$ arcsec the
contamination of random sources is high. Within 3 arcsec separation,
however, the fraction of false identifications is very low, $\lta
1$\%. This is lower than has been previously derived for simple
SDSS--FIRST comparisons (e.g. Ivezi{\'c} \etal\ 2002\nocite{ive02}, who
found a random contamination of about 9\% at 3 arcsec radius). This is
because the present analysis is limited to galaxies with NVSS counterparts
brighter than 5\,mJy.

Given the low contamination rate, all FIRST radio sources with offsets
below 3 arcsec can clearly be accepted as matches. If all sources between
3 and 10 arcsec were dropped, however, then the completeness would suffer.
Figure~\ref{firstoffsize} compares the offset of the FIRST sources against
their projected size along the offset direction, i.e. the product of the
deconvolved major axis length and the cosine of the angle between the
major axis and the offset vector. A significant fraction of the FIRST
sources with offsets between 3 and 10 arcsec are found to be extended
sources oriented close to the direction of the offset between the optical
and radio position. Only a few of the random positions are associated with
radio sources with these properties. The selection procedure was therefore
refined to accept those FIRST sources which are either (i) within 3
arcsec, or (ii) offset less than 10 arcsec, oriented within 30 degrees of
the offset vector, and offset by less than 75\% of the projected major
axis length of the source. These selection criteria are illustrated on
Figure~\ref{firstoffsize}. The addition of the 3--10 arcsec offset sources
significantly reduces the incompleteness of the sample for only a small
decrease in the reliability.

\begin{figure*}
\centerline{
\psfig{file=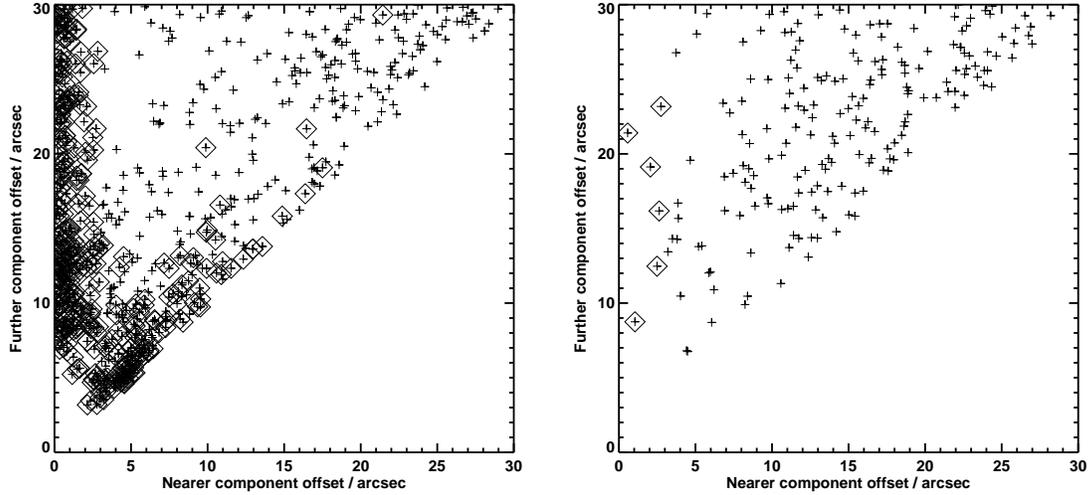,width=15cm,clip=}
}
\caption{\label{firstdoubles} For those SDSS galaxies with a single NVSS
match and two FIRST components associated with that, the left panel shows
the distribution of the positional offsets between the SDSS galaxies and
the two FIRST sources. The right panel shows the equivalent plot for
random positions. Accepted singles and doubles (defined by the criteria
described in the text) are indicated by diamonds.}
\end{figure*}

\begin{figure*}
\centerline{
\psfig{file=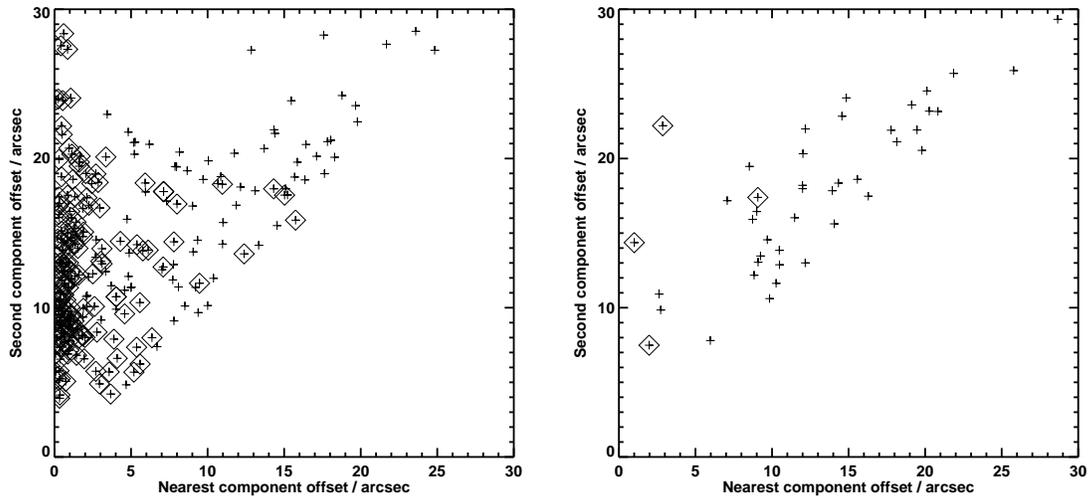,width=15cm,clip=}
}
\caption{\label{firsttriples} Left: the positional offsets between SDSS
galaxies and the nearest two FIRST sources, for those SDSS galaxies with a
single NVSS match within 30 arcsec, of $>5$ mJy flux density, and three FIRST
components associated with that. Right: the equivalent plot for random
positions. Accepted singles, doubles and triples (defined by the criteria
described in the text) are indicated.}
\end{figure*}

\subsubsection{Sources with two FIRST matches}

For galaxies with two FIRST matches, if the closer of the two matches is
within 3 arcsec then the source is accepted under the assumption that it
is either a single component source with a nearby unassociated source, or
the core of a core-jet source. There are 251 such SDSS galaxies (5.3
random). If neither FIRST source is within 3 arcsec, then it is possible
that the two FIRST components are two lobes of the same extended radio
source with no core.  As discussed earlier, McMahon \etal\
\shortcite{mcm02} found that in this case the two FIRST sources often have
the following properties: (i) they have comparable flux densities; (ii)
the flux-weighted mean position of the two sources is close to that of the
optical galaxy; (iii) the sizes of the two sources are comparable
(ie. both are lobes, not a core and a lobe). None of these conditions on
their own is sufficient to classify the source as a double without
including a lot of false detections, but the Monte--Carlo simulations show
that the combination of the three can be quite powerful. Galaxies with two
FIRST matches were accepted as double radio sources if the ratio of the
radio source sizes, multiplied by the ratio of the radio source flux
densities, multiplied by the offset in arcsec of the flux--weighted mean
position, is less than 5.  Figure~\ref{firstdoubles} shows the result of
this analysis: 116 double sources are selected among SDSS galaxies, but
only 1.2 for random positions, implying that the reliability is about 99\%.

It should be noted here that in cases where the galaxy is associated with
a single FIRST component, and the other component (or components, for the
cases with 3 or more matches discussed below) is genuinely unassociated,
the NVSS flux will overestimate the radio luminosity due to the
contaminating source.  However, only in very rare cases (ie. a faint point
source lying nearby a much brighter source) would such a correction be
significant and these cases could not reliably be separated from core--jet
type sources without visual analysis.

\subsubsection{Sources with three FIRST matches}

Galaxies for which there are three FIRST matches within 30 arcsec are
accepted if any of the following three conditions are satisfied: (i) the
nearest match is within 3 arcsec; (ii) any of the three pairs of sources
satisfies the criteria to be accepted as a FIRST double source; (iii) the
flux-weighted mean position of all three sources is within 3 arcsec of the
optical galaxy position, with the angle subtended by the outer two sources
relative to the middle one larger than 135 degrees (ie.  the source looks
like a straight(ish) triple source).  Figure~\ref{firsttriples} shows the
results of this selection.

\subsubsection{Sources with four or more FIRST matches}

Automated classification of more than three sources cannot be carried out
in an efficient and reliable way. For galaxies with four or more matches,
the nearest three matches are analysed using the criteria for three-source
matches to test whether they may be classified as triples, doubles or
singles. All galaxies not accepted in this way are sent for visual
analysis (a total of 23 optical galaxies or $\sim 0.01$\% of the SDSS
sample).

\subsection{Repeated matches}

The final list of matches was examined to ensure that two different SDSS
galaxies were not associated with the same NVSS source. This occurred on
24 occasions and these cases were all examined visually. In two cases, two
SDSS galaxies were associated with the same NVSS source and there was no
FIRST counterpart. In a further 14 cases, two SDSS galaxies were
associated with the same NVSS source which had a single FIRST counterpart
which lay close enough to both galaxies. For these 16 objects, the nearer
galaxy was accepted as the true match and the other galaxy was removed
from the radio source catalogue.  There were a further 8 cases where it
was found that two galaxies matched the same NVSS source but had distinct
FIRST counterparts. In other words, both galaxies were genuine radio
sources, but at the lower resolution of NVSS they had been convolved
together. In these cases the flux density of the NVSS source was divided
between the two galaxies according to the ratio of their integrated FIRST
flux densities. If the galaxies still remained above the 5\,mJy flux
density limit, they were retained in the radio source catalogue.

\begin{table}
\begin{center}
\caption{\label{makecat}Results of the analysis: the number of SDSS
galaxies, and the number of random locations (scaled to the same total
number of sources) falling into each category.}
\begin{tabular}{cccc}
\hline
\multicolumn{2}{c}{Source type}     & SDSS gals& Random            \\
\hline
NVSS multiples & Accepted directly  &    60    &       0.3         \\
NVSS multiples & Visual analysis    &  (277)   &     (92.1)        \\
               & of which confirmed &   109    &                   \\
NVSS singles   & 0 FIRST matches    &   134    &       8.7         \\
               & 1 FIRST match      &  1805    &      12.3         \\
               & 2 FIRST matches    &   367    &       6.5         \\
               & 3 FIRST matches    &   163    &       1.4         \\
               & 4+, confirmed      &    57    &       0.9         \\
               & 4+, visual analysis&   (23)   &      (4.5)        \\
               & of which confirmed &    17    &                   \\
\hline
Total          & Confirmed          &  2712    &      30.1         \\
\hline
\end{tabular}
\end{center}
\end{table}

\subsection{Completeness and Reliability of the matching procedure}

The results of the cross-matching procedure are provided in
Table~\ref{makecat}.  This table gives the number of SDSS galaxies
accepted as radio sources compared to the number of cases accepted from
the same number of random positions, for each different radio source
type. It therefore provides a direct measure of the reliability of each of
the criteria defined above. Overall, assuming visual analysis to be 100\%
reliable, only 30.1 false identifications are expected amongst the final
sample of 2712 radio sources. This corresponds to an overall reliability
of 98.9\%. The most unreliable part of the sample selection is for NVSS
sources without a FIRST counterpart. Of these, 6\% will be false
identifications.  This is unavoidable. If sources with no FIRST
counterpart are excluded, this reduces the completeness and strongly
biases the derived radio source sample by removing 5\% of the more
extended sources.

The completeness of the sample is more difficult to estimate than the
reliability, since the true number of matches expected is
unknown. However, various estimates can be made. For galaxies with
multiple NVSS components, a comparison of the number of candidate NVSS
doubles in the SDSS and random samples with the numbers accepted suggests
that the completeness is close to 90\%. For the single NVSS component
sources, Figure~\ref{nvssoffs} shows that there were 2973 SDSS galaxies
with an NVSS source within 15 arcsec, compared to only 311 random
galaxies.  Assuming that this excess is entirely due to genuine sources
and that all true matches lie within 15 arcsec, 2662 genuine
single--component NVSS sources are expected. Table~\ref{makecat} indicates
that 2543 single component sources were actually found by the adopted
selection procedures, of which about 30 will be false detections. An
estimate of the completeness is then (2543--30)\,/\,2662\,$=$\,94.4\%.
Note that this value is conservative because a fraction of the excess
matches are likely to be associated with companion galaxies, and so 2662
is an overestimate of the true number of expected matches. Therefore, the
overall completeness of the sample likely exceeds 95\%.

\begin{figure}
\centerline{
\psfig{file=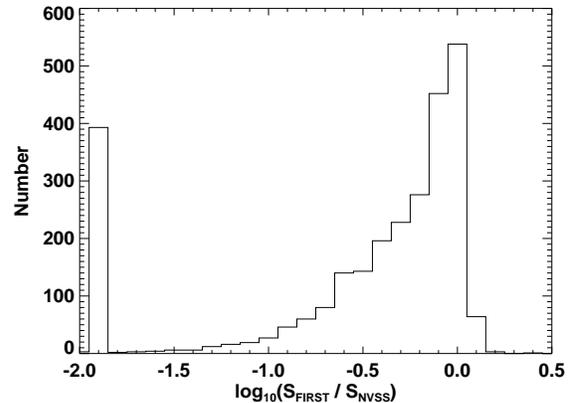,angle=90,width=8cm,clip=}
}
\caption{\label{fluxrat} A histogram of the FIRST to NVSS flux density
ratios of the final radio source sample. Sources without a detected
central FIRST component (either because they are resolved out, or because
they are extended multi--component FIRST sources) are included in the
left--most bin.}
\end{figure}

The values quoted for completeness and reliability are for all types of
radio source. There will be a small (but unavoidable) bias against
extended sources: the completeness for the single--component FIRST sources
approaches 100\%, while that of multi--NVSS--component sources is around
90\%.  Note that completeness estimates from previous cross--correlations
with the FIRST catalogue have not taken into account the sources missed
because sources with radio--optical offsets greater than 3 arcsec are
excluded ($\sim 3$\% of our final source catalogue) as are sources that
are completely resolved out by FIRST (5\%). These samples will also miss a
fraction of the extended NVSS sources (6\%) and the multi--component FIRST
sources (6\%). These omissions have a severe effect on the completeness of
any radio sample derived for the SDSS spectroscopic sample using the FIRST
survey alone. The radio luminosities of many sources would also be
underestimated using FIRST alone: the distribution of FIRST to NVSS flux
density ratios for the final sample of sources is plotted in
Figure~\ref{fluxrat}, and shows a long tail to low values. Note, however,
that all of these effects are somewhat less important when dealing with
the complete imaging catalogue of SDSS, for which the galaxies typically
lie at higher redshifts.

\subsection{The final radio source sample}

Details of the final SDSS radio sample of 2712 sources are given in
Table~\ref{cattable}.  This table provides the identification details of
each source so that they can be matched against either the original
spectra or against the catalogues of derived optical properties released
by Brinchmann \etal\ \shortcite{bri04b}. Also provided are the RA and Dec
of each source, the host galaxy redshift, the integrated NVSS flux density
and, where there is a central FIRST counterpart, the integrated flux
density, radio size and offset from the optical galaxy of the central
FIRST component. Each radio source is also given a classification to
identify its radio properties. Class 1 sources are single--component NVSS
sources with a single FIRST counterpart. Class 2 sources have a single
NVSS match which is resolved into multiple components by FIRST.  Class 3
sources have a single--component NVSS source, but no FIRST
counterpart. Class 4 sources have multiple NVSS components. The final
column of the table classifies each radio source as a star--forming galaxy
or a radio--loud AGN, according to the criteria described in
Section~\ref{radloudagn}.

\begin{table*}
\begin{center}
\caption{\label{cattable}Properties of the 2712 SDSS radio galaxies. Only
the first 30 sources are listed here: the full table is available
electronically. The first three columns give the identification of the
targetted galaxies through their plate and fibre IDs and the date of the
observations. Columns 4 to 6 give the RA, Dec and redshift of the
galaxies. Column 7 gives the integrated flux density of the source as
measured using the NVSS. Column 8 provides the radio classification of the
source: class 1 are single--component NVSS sources with a single FIRST
match; class 2 are single--component NVSS sources resolved into multiple
components by FIRST; class 3 are single--component NVSS source without a
FIRST counterpart; class 4 sources are those which have multiple NVSS
components. Where a galaxy has a central FIRST component, the integrated
flux density, offset from the optical galaxy, and radio size of that
central FIRST component are given in columns 9 to 11. The final column
provides the classification of the source as either a radio--loud AGN or a
star-forming galaxy, according to the criteria described in
Section~\ref{radloudagn}. Other properties of the host galaxy, such as
stellar mass, [OIII] emission line luminosity, $D_n(4000)$, etc can be
obtained from the data released by Brinchmann \etal\ (2004a), by
cross-comparing the host galaxy identifiers.}
\begin{tabular}{cccccccccccl}
\hline
Plate & Julian & Fibre &      RA      &      Dec     & z &
$S_{\rm NVSS}$ & Radio & $S_{\rm FIRST}$ & Offset & Radio & Source type \\

 ID   &  Date  &  ID   & \multicolumn{2}{c}{(J2000)} &   &
1.4\,GHz       & Class &  1.4\,GHz       &        & Size  &             \\

      &        &       &    (deg)     &    (deg)     &   & 
(Jy)           &       &    (Jy)         & (arcsec) & (arcsec) & \\
\hline
266 & 51630 &  25 & 146.95610 & -0.3423 & 0.1347 & 0.0963 & 1 & 0.1010 & 1.26 &  4.64 &   Radio-loud AGN \\
266 & 51630 &  90 & 146.14360 & -0.7416 & 0.2039 & 0.0068 & 1 & 0.0025 & 0.46 &  5.88 &   Radio-loud AGN \\
266 & 51630 & 119 & 146.73711 & -0.2522 & 0.1305 & 0.0075 & 1 & 0.0043 & 0.51 &  2.39 &   Radio-loud AGN \\
266 & 51630 & 141 & 146.37379 & -0.3684 & 0.0529 & 0.0104 & 1 & 0.0010 & 2.02 &  0.27 &   Star-forming \\
266 & 51630 & 223 & 145.60120 & -0.0014 & 0.1459 & 0.0054 & 1 & 0.0049 & 0.44 &  0.00 &   Star-forming \\
266 & 51630 & 506 & 146.46300 &  0.6387 & 0.0303 & 0.0052 & 1 & 0.0028 & 2.10 &  6.24 &   Star-forming \\
266 & 51630 & 517 & 146.37360 &  0.2555 & 0.1292 & 0.0275 & 1 & 0.0269 & 0.33 &  1.77 &   Radio-loud AGN \\
266 & 51630 & 543 & 146.80679 &  0.6656 & 0.0201 & 0.0180 & 1 & 0.0131 & 0.78 & 10.89 &   Radio-loud AGN \\
266 & 51630 & 545 & 146.79910 &  0.7027 & 0.0305 & 0.0063 & 1 & 0.0045 & 2.73 & 13.71 &   Star-forming \\
266 & 51630 & 572 & 146.78149 &  0.7380 & 0.2618 & 0.0489 & 1 & 0.0094 & 0.77 &  5.71 &   Radio-loud AGN \\
266 & 51630 & 613 & 147.08051 &  0.7880 & 0.2111 & 0.0081 & 1 & 0.0078 & 0.68 &  0.00 &   Radio-loud AGN \\
267 & 51608 &  34 & 149.16991 & -0.0233 & 0.1392 & 0.1661 & 4 & 0.0022 & 0.27 &  4.37 &   Radio-loud AGN \\
267 & 51608 &  47 & 148.43250 & -1.0264 & 0.1103 & 0.0092 & 1 & 0.0111 & 0.06 &  1.45 &   Star-forming \\
267 & 51608 &  97 & 148.23770 & -0.7920 & 0.0898 & 0.0207 & 1 & 0.0169 & 0.43 &  2.69 &   Radio-loud AGN \\
267 & 51608 & 205 & 147.70860 & -0.8878 & 0.2715 & 0.0558 & 1 & 0.0151 & 0.67 &  6.88 &   Radio-loud AGN \\
267 & 51608 & 260 & 147.42830 & -0.8401 & 0.0809 & 0.1671 & 1 & 0.0115 & 0.40 &  0.00 &   Radio-loud AGN \\
267 & 51608 & 297 & 147.20770 & -1.1859 & 0.1282 & 0.0129 & 1 & 0.0124 & 0.07 &  1.26 &   Radio-loud AGN \\
267 & 51608 & 497 & 148.23340 &  0.3577 & 0.2550 & 0.0570 & 1 & 0.0208 & 2.26 &  8.44 &   Radio-loud AGN \\
267 & 51608 & 512 & 148.37680 &  0.4487 & 0.0797 & 0.0118 & 1 & 0.0128 & 0.22 &  2.99 &   Radio-loud AGN \\
268 & 51633 &  38 & 150.47060 & -0.1252 & 0.0329 & 0.0364 & 1 & 0.0260 & 0.45 &  3.68 &   Radio-loud AGN \\
268 & 51633 & 271 & 149.28870 &  0.0092 & 0.1255 & 0.0108 & 1 & 0.0049 & 0.69 &  4.57 &   Radio-loud AGN \\
268 & 51633 & 394 & 148.95151 &  0.5569 & 0.0800 & 0.0117 & 1 & 0.0085 & 0.05 &  1.37 &   Radio-loud AGN \\
268 & 51633 & 433 & 149.33859 &  0.7065 & 0.0874 & 0.0055 & 1 & 0.0038 & 1.39 &  2.58 &   Star-forming \\
268 & 51633 & 461 & 149.35471 &  0.7309 & 0.0869 & 0.0080 & 1 & 0.0066 & 1.24 &  5.78 &   Star-forming \\
268 & 51633 & 479 & 149.39481 &  0.2386 & 0.1602 & 0.0229 & 1 & 0.0092 & 0.64 &  2.67 &   Radio-loud AGN \\
268 & 51633 & 489 & 149.64740 &  0.7428 & 0.0648 & 0.0065 & 1 & 0.0064 & 0.28 &  2.27 &   Star-forming \\
269 & 51910 &  59 & 151.36320 & -0.8746 & 0.2057 & 0.0051 & 1 & 0.0010 & 0.60 &  0.00 &   Radio-loud AGN \\
269 & 51910 & 171 & 151.33971 &  0.0886 & 0.1793 & 0.0087 & 1 & 0.0080 & 0.38 &  0.00 &   Radio-loud AGN \\
269 & 51910 & 257 & 150.47040 & -0.8781 & 0.1364 & 0.1507 & 4 &  ---   & ---  &  ---  &   Radio-loud AGN \\
269 & 51910 & 289 & 149.91280 & -1.2480 & 0.1375 & 0.0114 & 1 & 0.0111 & 0.29 &  1.39 &   Radio-loud AGN \\
\dots & \dots & \dots & \dots & \dots & \dots & \dots & \dots & \dots & \dots & \dots &\multicolumn{1}{c}{\dots} \\
\hline
\end{tabular}
\end{center}
\end{table*}

\section{Definition of the radio-loud AGN sample}
\label{radloudagn}

The sample of radio--emitting galaxies contains both radio--loud AGN and a
population of star forming galaxies. The latter emit at radio wavelengths
mostly as a result of the synchrotron emission of particles accelerated in
supernova shocks, and their radio luminosity is therefore roughly
correlated with their star formation rate: a 1.4\,GHz radio luminosity of
$10^{22}$\,W\,Hz$^{-1}$ corresponds to a star formation rate of order $5
M_{\odot}$\,yr$^{-1}$ (e.g. Condon 1992 and references therein; Carilli
2001)\nocite{con92,car01b}. In order to investigate the host galaxies of
these two populations, it is first necessary to separate the radio--loud
AGN from the star--forming galaxies.

Star--forming galaxies and AGN are often separated using optical
emission--line properties. Sadler \etal\ \shortcite{sad02} used a visual
emission--line classification in their study of radio sources in the
2dFGRS: radio--emitting galaxies without detectable emission lines were
classified as radio--loud AGN.  Kauffmann \etal\ \shortcite{kau03c} used
the location of a galaxy in the [OIII]~5007 / H$\beta$ versus [NII]~6583 /
H$\alpha$ emission line diagnostic diagram (Baldwin, Phillips \& Terlevich
1981; hereafter BPT)\nocite{bal81} to separate optical AGN from normal
star forming galaxies. A key result of the Kauffmann \etal\ study was that
a significant fraction of emission-line selected AGN also have associated
star formation. This result means that optical line ratio diagnostics
should not be used to identify radio--loud AGN, because star formation
activity in galaxies with a radio--quiet active nucleus would give rise to
radio emission (and hence a radio--loud classification). In addition, for
galaxies which do contain a genuine radio--loud AGN, the radio luminosity
associated with the active nucleus will be overestimated if there is a
significant contribution of star formation to the radio emission.

\begin{figure*}
\centerline{
\psfig{file=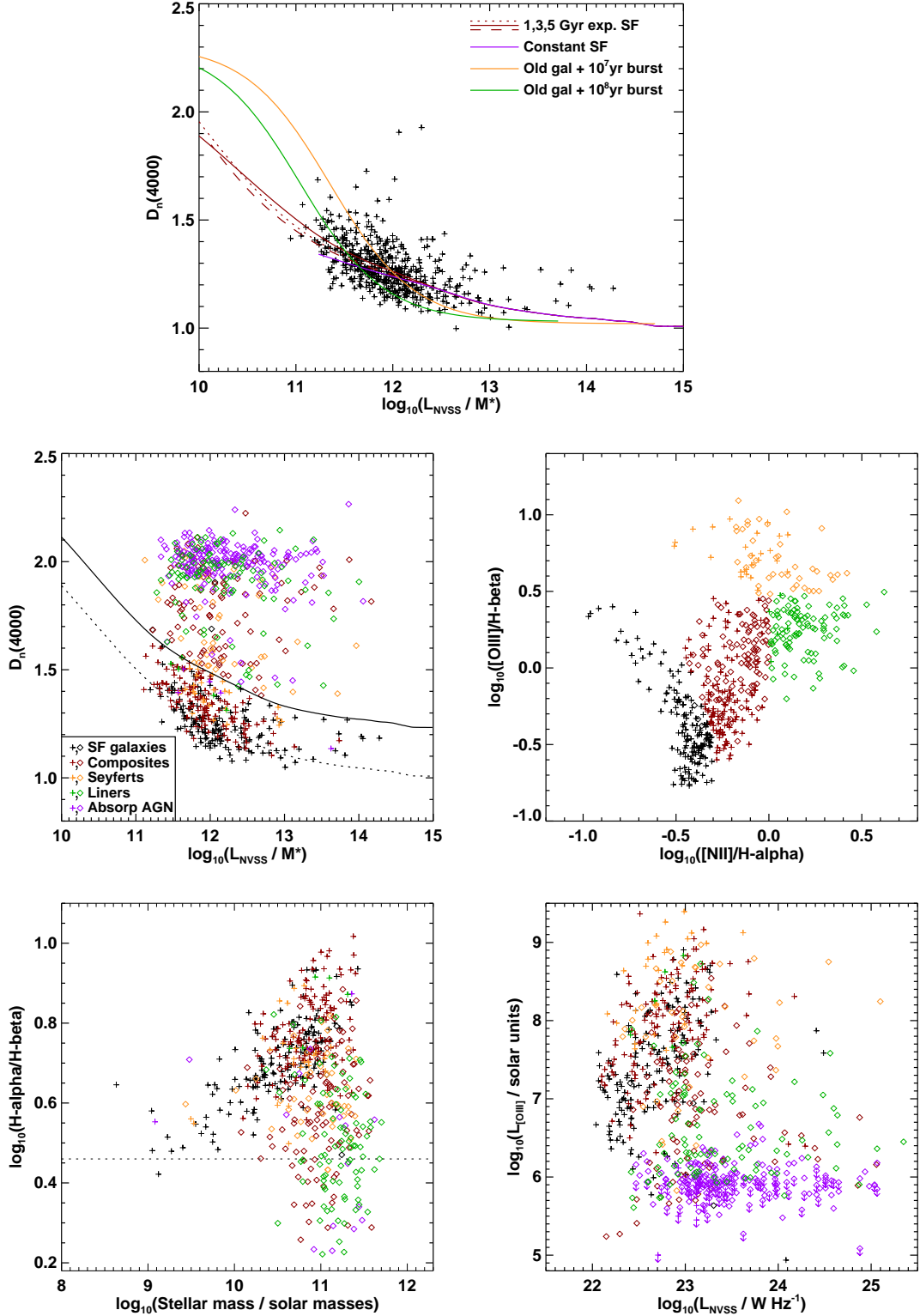,angle=0,width=15cm,clip=}
}
\caption{\label{sfagncut} The separation of radio--loud AGN from
starbursting galaxies in SDSS. The {\bf top} panel shows $D_n(4000)$
versus $L_{\rm 1.4GHz} / M_*$ for radio--emitting galaxies (brighter than
2.5\,mJy) classified as star forming in the [OIII]~5007 / H$\beta$ versus
[NII]~6583 / H$\alpha$ emission line diagnostic diagram using the criteria
of Kauffmann \etal\ (2003c). Overlaid on this are theoretical predictions
for galaxies with different star formation histories, as derived using the
Bruzual \& Charlot (2003) stellar synthesis models (see text for
details). These models are able to explain the locus of the star forming
galaxies. The {\bf middle left} panel shows this same plot for all
radio--emitting galaxies brighter than 5mJy. The different colours
represent different galaxy classifications based upon their locations in
the emission line diagnostic diagram. The dotted line is the 3\,Gyr
exponentially decaying star formation track of the top panel and the solid
line, 0.225 above this in $D_n(4000)$, is the proposed division between
radio-loud AGN (indicated by diamonds) and galaxies whose radio emission
is dominated by star formation (crosses). The colours and symbols are
identical on the following three panels.  The {\bf middle right} panel
shows the emission line diagnostic diagram for these galaxies, the {\bf
lower left} panel shows the H$\alpha$\,/\,H$\beta$ ratio versus stellar
mass (the dotted line indicates the value for zero reddening), and the
{\bf lower right} panel shows the distribution of the galaxies in the
$L_{\rm [OIII]~5007}$ versus $L_{\rm 1.4 GHz}$ plane. These final three
plots show that the AGN--starburst classification derived works very
well.}
\end{figure*}

Machalski \& Condon \shortcite{mac99} studied radio galaxies in the LCRS
and used far--infrared to radio flux density ratios and far--infrared
spectral indices to separate the radio--loud AGN and star forming
populations. The far--infrared radio correlation for star--forming
galaxies (e.g. Yun, Reddy \& Condon 2001)\nocite{yun01} could also be used
to correct for the contribution of star formation to the radio
luminosities of these systems.  This is perhaps the ideal method, but
unfortunately the Infrared Astronomical Satellite (IRAS) Faint Source
Catalogue is not quite deep enough to allow this to be used for the SDSS
galaxies in this paper\footnote{In fact, the IRAS Faint Source Catalogue
is deep enough that the {\it majority} of star--forming galaxies with
1.4\,GHz radio flux densities of 5\,mJy are detected (and lie as expected
on the far--infrared radio correlation), plus a few of the AGN, but the
observations are not deep enough that {\it all} star forming galaxies are
detected. The IRAS data cannot therefore be used as a discriminant between
the two subclasses, nor to correct for any star formation contribution to
the radio emission of the AGN.}. A variety of alternative methods were
therefore considered, and a procedure based on the location of a galaxy in
the plane of $D_n(4000)$ versus $L_{\rm 1.4GHz} / M_*$ was adopted. The
$L_{\rm 1.4GHz} / M_*$ ratio provides the radio luminosity per stellar
mass of the galaxy and $D_n(4000)$ is a fairly accurate indicator of mean
stellar age for ages below about a Gyr (at higher ages it is also
sensitive to metallicity; cf. Kauffmann \etal\ 2003b).\nocite{kau03a}
Thus, star forming galaxies would be expected to occupy a well--defined
locus in this plane, while radio--loud AGN would be offset to higher radio
luminosities. This is demonstrated in the first two panels of
Figure~\ref{sfagncut}.

\nocite{kau03c,bru03}

The top panel of Figure~\ref{sfagncut} shows $D_n(4000)$ versus $L_{\rm
1.4GHz} / M_*$ for radio--emitting galaxies that are classified as star
forming galaxies using the [OIII]~5007 / H$\beta$ versus [NII]~6583 /
H$\alpha$ emission line diagnostic diagram.  The criteria of Kauffmann
\etal\ \shortcite{kau03c} have been adopted and only galaxies with
redshifts in the range $0.03 \le z \le 0.10$ have been plotted; the lowest
redshifts are excluded because aperture corrections are substantial,
whilst beyond $z=0.1$ the sensitivity to emission lines is low, hampering
classification by emission--line diagnostics.  Radio luminosities have
been calculated from the fluxes assuming a radio spectral index of 0.7.

Overlaid on this are theoretical predictions, derived using the Bruzual \&
Charlot \shortcite{bru03} stellar synthesis models, for the location of
galaxies with different star formation histories. For these models, the
radio luminosities have been calculated using the prescription of Hopkins
\etal\ \shortcite{hop01}: $L_{\rm 1.4GHz} = 1.8 \times 10^{21} ({\rm
SFR}/M_{\odot})$W\,Hz$^{-1}$, where SFR is the average star formation rate
in the past $10^8$ years. Three of the models are for galaxies with
exponentially--decaying star formation rates, of characteristic timescales
1,3 and 5 Gyrs; the tracks indicate how these galaxies move across this
plane as they age. A fourth model shows the track for a galaxy with a
constant star formation rate.  Two further models consider an old
(10\,Gyr) galaxy which has undergone a recent burst of star formation
($10^7$ and $10^8$ years ago), as might be the case for a merger-triggered
event. Here, the loci of the tracks show what happens if different
fractions of the total galaxy mass are converted into stars in the
burst. These theoretical tracks largely cover the location of the data
points.

The middle left panel of Figure~\ref{sfagncut} shows the same plot, but
now includes all radio--emitting galaxies in this redshift range.  The
different colours represent different galaxy classifications based upon
their locations in the BPT diagram (black -- star forming galaxy; red --
composite systems with both star formation and an AGN; orange -- Seyfert
AGN; green -- LINER AGN; purple -- no emission lines). A significant
number of the objects classified as AGN based on their emission--line
ratios overlap the star--forming population in this plane. This is
interpreted as meaning that these are radio--quiet AGN whose radio
emission is due to star formation -- the problem identified above.  The
dotted line on this plot shows the 3\,Gyr exponential star formation model
from the top panel. The solid line, 0.225 above this in $D_n(4000)$, is
the proposed division between radio-loud AGN (above the line, plotted as
diamonds) and star--forming galaxies (below the line, plotted as
crosses). This cut--off value was chosen to be most consistent with other
methods that could have been adopted for AGN--starburst separation, as
illustrated in the later panels. Using this cut-off, 2215 radio sources
are classified as radio--loud AGN, and 497 as star--forming. Note that the
plots only show the subset of these at redshifts with $0.03 \le z \le
0.1$, to avoid overcrowding the plot and to allow a comparison with
emission--line diagnostic methods; the higher redshift sources fill out
more of the plane at larger values of $L_{\rm 1.4GHz} / M_*$, and confirm
that the location of the proposed cut at those values is appropriate.

The middle right panel shows the BPT emission line diagnostic diagram. It
can be seen that the AGN--starburst separation defined above (ie. diamonds
versus crosses) also makes good sense in this plot: (i) the `composite'
galaxies that lie close to the star forming galaxy locus are largely
classified as starbursts, whilst those nearer to the AGN locus are
predominantly classified as radio--loud AGN; (ii) almost all of the LINERS
are classified as radio--loud AGN; (iii) the Seyferts close to the LINER
region are mostly classified as radio-loud, whilst those with lower
[NII]~6583 / H$\alpha$ ratios are a mixture of the two classes; (iv) the
three star forming galaxies now classified as radio-loud AGN all lie near
the boundary with composites.

The lower left panel shows the H$\alpha$ / H$\beta$ line ratio as a
function of galaxy stellar mass.  The H$\alpha$ / H$\beta$ line ratio is an
approximate measure of dust--reddening; the dotted line shows the expected
value for zero reddening.  Star forming galaxies form a tight relation
between these parameters, with more massive galaxies being more heavily
reddened (cf. Figure 6 of Brinchmann \etal\ 2004b).\nocite{bri04a}
Radio--loud AGN deviate from this locus, in the sense of having less
reddening (due to less star formation and hence less dust) at a given
stellar mass; this diagram indicates that the classification division
adopted for $D_n(4000)$ versus $L_{\rm 1.4GHz} / M_*$ works well.

The final panel shows the distribution of the galaxies in the $L_{\rm
[OIII]~5007}$ versus $L_{\rm 1.4 GHz}$ plane. This relation was considered
as a way to separate radio--loud AGN and star-forming galaxies; indeed, it
can be seen that for the LINERS and the galaxies without emission lines
the division agrees very well with that adopted. However, many of the
Seyferts and composites lie on the relation defined by the star--forming
galaxies in this plane, but are considerably offset from the star forming
locus in all of the other plots. It is for this reason that the final
classification was not based upon this relation.

These plots demonstrate the reliability of the AGN--starburst separation
using the $D_n(4000)$ versus $L_{\rm 1.4GHz} / M_*$ relation: through
comparison of the locations of galaxies on different diagnostics, it is
estimated that $\lta 1$\% of objects will have been misclassified.  The
$D_n(4000)$ versus $L_{\rm 1.4GHz} / M_*$ relation was also used to
estimate and to correct for the star formation contribution to the radio
luminosity of galaxies classified as radio--loud AGN: for each of these
galaxies the `star formation' radio luminosity corresponding its 4000\AA\
break strength, as estimated by the 3\,Gyr exponential star formation
track (the dotted line in the middle left panel), was subtracted to obtain
a corrected AGN radio luminosity. In no case was this correction larger
than 15\%.

\begin{figure}
\centerline{
\psfig{file=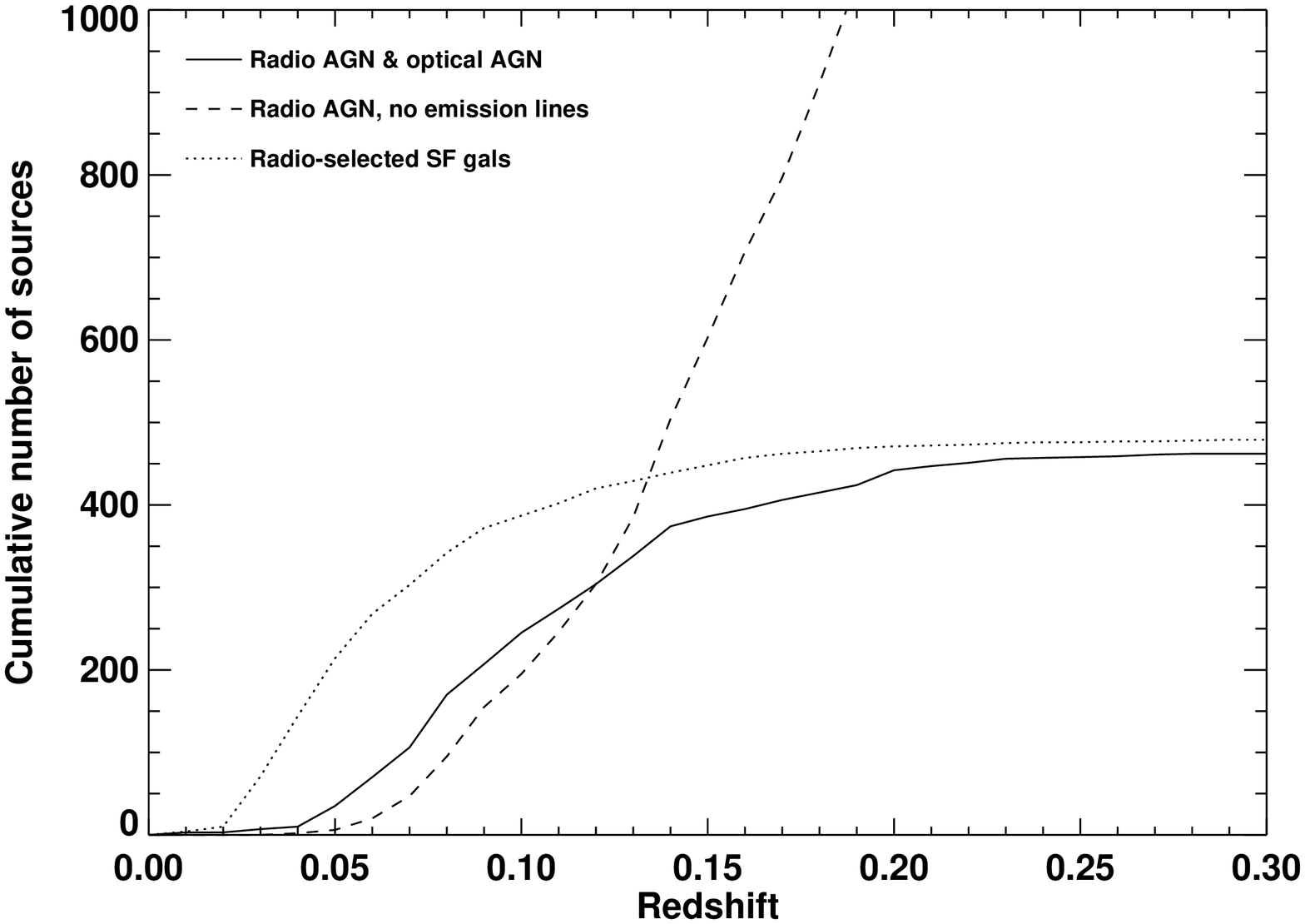,angle=90,width=8.6cm,clip=}
}
\caption{\label{cumfracs} The cumulative number, as a function of
redshift, of radio--selected star forming galaxies, optically--passive
radio--loud AGN and radio--loud AGN with optical AGN characteristics, in
the SDSS sample. [The line for AGN with emission lines continues to rise
to nearly 1800 by $z=0.3$].}
\end{figure}

The radio--loud AGN in the sample exhibit a variety of optical properties;
some are classified as optical AGN based upon their emission lines while
others are optically inactive. Figure~\ref{cumfracs} shows the cumulative
fractions of the different radio source types as a function of
redshift. Out to redshifts $z \sim 0.1$, the relative numbers of
radio--loud AGN with and without emission lines are roughly similar. At
higher redshifts the proportion of emission-line AGN decreases rapidly;
this is because emission lines such as [OIII]~5007 become increasing
difficult to detect at higher redshift (only lines brighter than $\sim
10^{5.8} L_{\odot}$ can be detected at $z =0.1$), both because of the
increased distance and because the larger physical size of the
spectroscopic fibres means that a larger fraction of starlight from the
host galaxy is included. This makes it more difficult to pick out the
weaker nuclear lines.

\begin{figure*}
\centerline{
\psfig{file=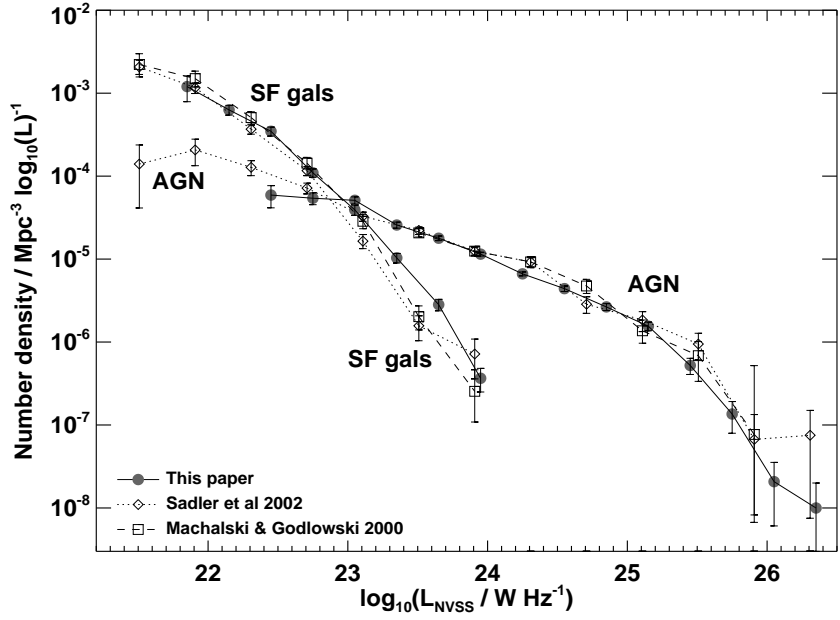,angle=90,width=12cm,clip=}
}
\caption{\label{lumfuncplot} The local radio luminosity function at
1.4\,GHz derived separately for radio--loud AGN and star--forming
galaxies. Filled points connected by solid lines indicate the data derived
in this paper, using the SDSS. No correction has been made for
incompleteness or misidentification, but such effects are small. For
comparison, the results of Sadler \etal\ (2002) using the 2dFGRS, and
those of Machalski \& Godlowski (2000) using the LCRS are also
plotted. The AGN results are in excellent agreement between the different
samples. The results for star--forming galaxies are in broad agreement but
with small differences in normalisation at the larger luminosities.}
\end{figure*}

\begin{table*}
\caption{\label{lumfunctab} The local radio luminosity function at
1.4\,GHz, derived separately for radio--loud AGN and star--forming
galaxies.}
\begin{tabular}{rrrrrrr}
\hline
&
\multicolumn{2}{c}{All radio sources} &
\multicolumn{2}{c}{Radio--loud AGN} &
\multicolumn{2}{c}{Star forming galaxies} \\

${\rm log}_{10} L_{\rm 1.4 GHz}$ &
N & ${\rm log}_{10} \rho$~~~~~~~~~~ &
N & ${\rm log}_{10} \rho$~~~~~~~~~~ &
N & ${\rm log}_{10} \rho$~~~~~~~~~~ \\

(W Hz$^{-1}$) & 
& (log$_{10}L)^{-1}$Mpc$^{-3}$ &
& (log$_{10}L)^{-1}$Mpc$^{-3}$ &
& (log$_{10}L)^{-1}$Mpc$^{-3}$ \\
\hline
21.60-21.90~~ &  10 & $-2.92^{+0.13}_{-0.18}$~~ &     &                           & 10 & $-2.92^{+0.13}_{-0.18}$~~ \\
21.90-22.20~~ &  64 & $-3.19^{+0.05}_{-0.06}$~~ &     &                           & 64 & $-3.20^{+0.05}_{-0.06}$~~ \\
22.20-22.50~~ & 105 & $-3.39^{+0.05}_{-0.05}$~~ &  15 & $-4.23^{+0.11}_{-0.15}$~~ & 90 & $-3.46^{+0.05}_{-0.06}$~~ \\
22.50-22.80~~ & 132 & $-3.79^{+0.04}_{-0.04}$~~ &  43 & $-4.27^{+0.07}_{-0.08}$~~ & 89 & $-3.96^{+0.05}_{-0.05}$~~ \\
22.80-23.10~~ & 195 & $-4.04^{+0.03}_{-0.04}$~~ & 115 & $-4.29^{+0.04}_{-0.05}$~~ & 80 & $-4.41^{+0.06}_{-0.07}$~~ \\
23.10-23.40~~ & 219 & $-4.44^{+0.03}_{-0.03}$~~ & 157 & $-4.59^{+0.03}_{-0.04}$~~ & 62 & $-4.99^{+0.05}_{-0.06}$~~ \\
23.40-23.70~~ & 335 & $-4.68^{+0.02}_{-0.03}$~~ & 291 & $-4.75^{+0.03}_{-0.03}$~~ & 44 & $-5.55^{+0.06}_{-0.07}$~~ \\
23.70-24.00~~ & 459 & $-4.93^{+0.02}_{-0.02}$~~ & 448 & $-4.94^{+0.02}_{-0.02}$~~ & 11 & $-6.44^{+0.12}_{-0.17}$~~ \\
24.00-24.30~~ & 389 & $-5.12^{+0.05}_{-0.05}$~~ & 389 & $-5.12^{+0.02}_{-0.02}$~~ &    & \\
24.30-24.60~~ & 303 & $-5.33^{+0.03}_{-0.04}$~~ & 303 & $-5.33^{+0.03}_{-0.03}$~~ &    & \\
24.60-24.90~~ & 184 & $-5.56^{+0.04}_{-0.04}$~~ & 181 & $-5.56^{+0.04}_{-0.04}$~~ &    & \\
24.90-25.20~~ & 108 & $-5.81^{+0.05}_{-0.05}$~~ & 108 & $-5.81^{+0.05}_{-0.05}$~~ &    & \\
25.20-25.50~~ &  35 & $-6.28^{+0.09}_{-0.11}$~~ &  35 & $-6.28^{+0.09}_{-0.11}$~~ &    & \\
25.50-25.80~~ &   8 & $-6.87^{+0.15}_{-0.23}$~~ &   8 & $-6.87^{+0.15}_{-0.23}$~~ &    & \\
25.80-26.10~~ &   2 & $-7.68^{+0.23}_{-0.53}$~~ &   2 & $-7.68^{+0.23}_{-0.53}$~~ &    & \\
26.10-26.40~~ &   1 & $-8.00^{+0.30}_{-1.00}$~~ &   1 & $-8.00^{+0.30}_{-1.00}$~~ &    & \\
\hline
\end{tabular}
\end{table*}

\section{The local radio luminosity function}
\label{radlumfunc}

The local radio luminosity function was derived both for radio--loud AGN
and radio--emitting star--forming galaxies out to redshift 0.3. These were
calculated in the standard way using the $1/V_{\rm max}$ method
\cite{sch68b,con89}, where $V_{\rm max}$ was calculated using the upper
and lower redshift limits determined by the joint radio and optical
selection criteria, namely a radio cut--off of 5\,mJy and optical
cut--offs of $14.5 < r < 17.77$. An accurate calculation of the exact area
of sky within the overlap region of the SDSS DR2 and the FIRST survey is
not simple.  The absolute normalisation of the derived radio luminosity
function has thus been set by normalising the total radio luminosity
function to match that derived for the 2dFGRS by Sadler \etal\
\shortcite{sad02} over the radio luminosity range $10^{23}$ to
$10^{24.5}$W\,Hz$^{-1}$, where the errors on the two luminosity function
determinations are both small. Note that no correction has been made for
incompleteness or misidentification in the radio samples but, as discussed
above, it is expected that this will be relatively small.

The radio luminosity functions are tabulated in Table~\ref{lumfunctab}.
The uncertainties quoted on the luminosity function determination are the
statistical Poissonian errors only; these have become so small for the
SDSS sample at some luminosities that systematic errors are likely to
dominate. One important source of systematic error will be cosmic
variance. Another is the separation of AGN and star forming galaxies: for
the highest luminosity bin of the star forming galaxies and the lowest
luminosity bin of the AGN, this systematic error is likely to be
comparable to or larger than the Poissonian uncertainties.

The radio luminosity functions are displayed in Figure~\ref{lumfuncplot},
along with the results of Sadler \etal\ \shortcite{sad02} for the 2dFGRS
and Machalski \& Godlowski (2000) for the LCRS (both corrected to the
cosmology adopted in this paper; note that these determinations are not
corrected for incompleteness either). The luminosity function of
radio--loud AGN generally agrees well with these previous analyses,
although with notably smaller errors. The apparent small mismatches
between the Sadler \etal\ results and those of this paper at $10^{24.5}$
and $10^{25.5}$W\,Hz$^{-1}$ are likely to be due to cosmic variance.

The luminosity function of star forming galaxies has similar shape to
previous measurements, but with slightly higher space densities at high
luminosities ($L_{\rm 1.4GHz} > 10^{23}$\,W\,Hz$^{-1}$). There are three
possible differences between the analyses which could account for
this. First, the different optical magnitude limits of the different
surveys may influence the population of radio star--forming galaxies
studied. Second, the combined FIRST--NVSS radio--optical
cross--correlation method adopted here will lead to a more complete
sample, particularly for low-redshift star forming galaxies with extended
radio emission. Third, the disparity might arise from the contrasting ways
in which different analyses treat radio--quiet AGN with associated star
formation activity, for which the radio emission is due to the star
formation.  The technique used in the current paper for separating the
star forming and AGN populations would classify such objects as star
forming galaxies, but in emission line ratio classifications (as used, for
example, by Sadler \etal) they might be classified as AGN. This would lead
to previous studies estimating a lower space density of star forming
galaxies, particularly at the highest radio luminosities. Note that if
star formation dominates the radio and far--infrared emission of these
radio--quiet AGN, then these objects would be expected to lie on the
far--infrared radio correlation for star forming galaxies, despite being
classified (by emission line means) as AGN. In this respect it is
interesting to consider Figure~10 of Sadler \etal\, which shows the
far--infrared radio relation for the objects in their sample; many of the
objects which lie on the far-infrared radio relation for star forming
galaxies, but which have radio luminosities $L_{\rm 1.4GHz} >
10^{23}$\,W\,Hz$^{-1}$, are indeed classified as AGN. If just some of
these objects are truly radio quiet AGN, with radio emission due to
associated star formation activity, they could easily account for the
small difference between the luminosity function determinations.

It is instructive to compare the radio luminosity function for star
forming galaxies with that derived at far-infrared (FIR)
wavelengths. Star--forming galaxies show a tight correlation between their
radio and far--infrared luminosities, which Yun \etal\ \shortcite{yun01}
showed for the range of luminosities probed in the current study to be
indistinguishable from a linear relation: $L_{\rm 1.4GHz} / {\rm W
Hz}^{-1} = 10^{11.95} L_{{\rm 60}\mu{\rm m}} / L_{\odot}$. The local FIR
luminosity function need only be adjusted by this factor to estimate the
local radio luminosity function.

\begin{figure}
\centerline{
\psfig{file=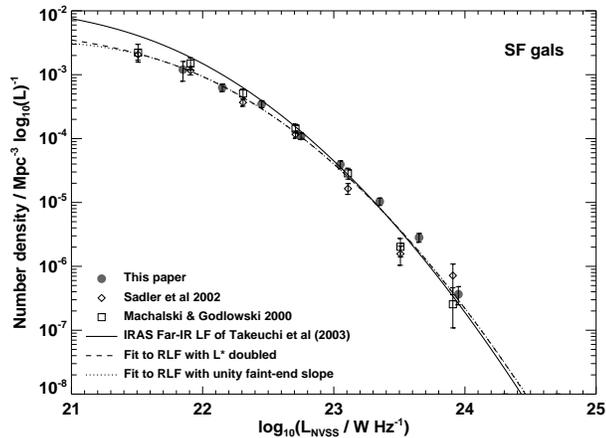,angle=90,width=8.5cm,clip=}
}
\caption{\label{sflumfunc} The local radio luminosity function at 1.4\,GHz
derived for star--forming galaxies; symbols are as in
Figure~\ref{lumfuncplot}. The solid line represents the far-IR luminosity
function as derived by Takeuchi \etal\ 2003 using the IRAS PSCz survey,
converted to the radio band using the radio to far--IR conversion of Yun
\etal\ 2001. The dotted and dashed lines show that improved fits at low
luminosities can be obtained by, respectively, doubling the characteristic
luminosity or setting the faint end slope to unity (with corresponding
changes in $\phi_*$).}
\end{figure}

Takeuchi \etal\ \shortcite{tak03} derived the local FIR luminosity
function using the IRAS Point Source Catalogue redshift survey (PSCz;
Saunders \etal\ 2000)\nocite{sau00}. They fitted the data with an analytic
function of the form suggested by Sandage \etal\ \shortcite{san79}, namely

\begin{displaymath}
\phi(L) = \phi_* \left(\frac{L}{L_*}\right)^{1 - \alpha} {\rm exp}
\left(\frac{-1}{2\sigma^2} \left[{\rm log}
\left(1+\frac{L}{L_*}\right)\right]^2 \right)
\end{displaymath}

\noindent with $\alpha = 1.23 \pm 0.04$, $\phi_* = (2.60 \pm 0.30) \times
10^{-2} h^3$Mpc$^{-3}$, $\sigma = 0.724 \pm 0.01$ and a characteristic
luminosity of $L_* = (4.34 \pm 0.87) \times 10^8 h^{-2} L_{\odot}$ (where
$h$ is $H_0$ in units of 100\,km\,s$^{-1}$Mpc$^{-1}$). Converting the
characteristic luminosity by the factor given above, the equivalent local
radio luminosity function is shown as the solid line in
Figure~\ref{sflumfunc}. This provides a good match to the data at high
radio luminosities, supporting the division adopted between AGN and star
forming galaxies. The match is less than perfect at lower luminosities,
however.  The luminosity range covered by the radio observations is not
sufficient constrain the parameters for a fit of the above form using the
radio data, but the dotted and dashed lines show the effect of doubling
the characteristic luminosity or setting the faint end slope $\alpha$ to
unity (with corresponding changes in $\phi_*$); these provide a much
better fit to the data. However, these represent much larger changes than
are allowed by the errors on the fitted FIR parameters or on the radio to
FIR conversion. 

The difference between the radio and FIR luminosity functions means that
the observed radio to FIR correlation stops being a linear relation at low
luminosities. This has been suggested before in terms of a steepening of
the relation below $L_{{\rm 60}\mu{\rm m}} \sim 10^9 L_{\odot}$ (cf. Yun
\etal\ 2001 and references therein) but questions have been raised as to
whether the previous studies may be affected by selection biases in the
samples under study. Here it is shown that the difference is also present
in the luminosity functions. The difference occurs for the lowest
luminosity sources, which are generally at the lowest redshifts and
therefore may have larger angular sizes, in which case it may be caused by
photometric errors due to aperture effects. This is unlikely to be the
case, however. The NVSS is sensitive to emission on angular scales out to
several arcmins, much larger than potential host galaxies. The IRAS
photometry is susceptible to missing extended emission, but for PSCz
galaxies this has been corrected for \cite{sau00}, and in any case any
missing IRAS flux would lower the far-IR to radio ratio, which is in the
opposite sense to the observed differences. More likely is that there is a
genuine effect at work: either the radio luminosity or the FIR luminosity
is not directly proportional to the star formation. One way in which this
might occur is if there is an additional low--level contribution to the
FIR luminosity of galaxies, for example from dust heated by low mass
stars, which is usually swamped by the FIR emission associated with star
formation, but becomes significant at low star formation rates
(cf. Devereux \& Eales 1989 and references therein).\nocite{dev89}

\section{Conclusions} 
\label{discuss}

The main results of this paper are as follows:

\begin{itemize}
\item A catalogue of 2712 radio sources has been derived by
cross--correlating the SDSS spectroscopic sample with a combination of the
NVSS and FIRST surveys.

\item The use of a hybrid NVSS--FIRST method to identify the radio sources
has been highly successful, resulting in a sample with a reliability of
98.9\% and a completeness which is estimated to be over 95\%.

\item The radio sources have been sub--divided into 2215 radio--loud AGN
and 497 star--forming galaxies, based upon their location in the plane of
4000\AA\ break strength versus radio luminosity per unit stellar mass.

\item The local radio luminosity functions of radio--loud AGN and star
forming galaxies have been derived separately. These are in excellent
agreement with previous studies, but with smaller uncertainties.

\item The local radio  luminosity function of star--forming galaxies 
has been compared to that derived in the
far--infrared. Differences between the two confirm that the far--IR to
radio correlation becomes non--linear at low luminosities.
\end{itemize}

The development of the hybrid NVSS--FIRST method for identification of
radio sources represents a large step forward in the study of radio source
host galaxies. One note of caution needs to be added: the parameters for
acceptance or rejection of sources have been optimised for the SDSS
spectroscopic sample, and if the comparison survey has a significantly
different sky surface density of objects, the offset parameters for
acceptance may need to be modified in order to retain optimal completeness
and reliability.  The general method, however, can (and where possible,
should) be adopted unchanged in future studies.

The sample of radio sources produced will prove invaluable in the study of
the host galaxies of radio--loud AGN, due to the large size of the sample
and the wealth of information available on the host galaxies from the
SDSS. Such analysis is the focus of the accompanying paper.

\section*{Acknowledgements} 

PNB would like to thank the Royal Society for generous financial support
through its University Research Fellowship scheme. The authors thank Jarle
Brinchmann, Stephane Charlot, Christy Tremonti and Simon White for making
their catalogues available and for useful discussions. The research makes
use of the SDSS Archive, funding for the creation and distribution of
which was provided by the Alfred P. Sloan Foundation, the Participating
Institutions, the National Aeronautics and Space Administration, the
National Science Foundation, the U.S. Department of Energy, the Japanese
Monbukagakusho, and the Max Planck Society.  The research uses the NVSS
and FIRST radio surveys, carried out using the National Radio Astronomy
Observatory Very Large Array: NRAO is operated by Associated Universities
Inc., under co-operative agreement with the National Science Foundation.

\bibliography{pnb} 
\bibliographystyle{mn} 

\label{lastpage}

\end{document}